# Science and survival: insights from Astronomy[1]

**Ewine F. van Dishoeck**, *Leiden Observatory, Leiden University, the Netherlands; President, International Astronomical Union*
**Debra M. Elmegreen**, *Vassar College, Poughkeepsie, USA; President-elect, International Astronomical Union*

**Introduction**

The COVID-19 pandemic has focused worldwide attention on the critical need for science in our lives. Medical doctors and health-related professionals are on the forefront in the global effort to combat the virus and its spread, but scientists in other fields also are contributing to counteract this devastating disease and its side effects on society. Astronomers are among them, and this brief paper outlines some of the lessons that come from studying our place in the Universe and the global coordination of scientific and outreach activities.

Exploring the wonders of the Universe and viewing the starry night sky are opportunities available to anyone, anywhere in the world. This accessibility allows astronomy to play a diverse role within society, from maintaining our shared sense of connectivity, to offering inspiration, to utilizing a rare combination of skills through STEM learning, technology development and research [1]. Astronomers, through the International Astronomical Union (IAU) and other international organizations, are strongly committed to sharing with society the many benefits that astronomy brings to education, employability and global citizenship. Here we highlight the diverse roles of astronomy in (i) sparking curiosity, especially in young children, which is essential for innovation; (ii) technological advances with societal spinoffs; (iii) education and outreach; (iv) using astronomy for development and capacity building; and (v) providing perspective on our Earth as viewed from space.

**World-wide coordination in astronomy**

Astronomy seeks to provide insight into the origin and evolution of stars, planets, galaxies and even the Universe itself. Where do we come from? Are we alone? What is the future of our Sun and its solar system, and of the Milky Way, the galaxy of which we are part? These are some of the biggest questions that humankind can ask, appealing to deep cultural and philosophical yearnings. They are also the questions that fascinate the public at large, young and old. To answer these questions, astronomers need increasingly larger and more sophisticated telescopes, in space and on the ground, that can only be funded and realized by large international organizations and through collaborations that now span the entire globe. The Vatican Observatory (Specola Vaticana) , with its telescopes originally near Rome (Castel Gandolfo) and now in Arizona, has been part of these global collaborations from the very start, beginning with the Carte du Ciel in the late 19[th] century. The massive amounts of data accumulated by modern facilities are preserved in huge archives, most of which are made public to the entire world: anyone with internet access can download them. Thus, the astronomical community has a tradition of working in large, multicultural collaborations that openly share data to advance science.

---

Astronomers are coordinated through the International Astronomical Union (IAU) that brings together more than 14 000 professional astronomers from more than 100 countries worldwide. Its mission is to promote and safeguard astronomy in all its aspects through international cooperation. Founded in 1919, the IAU was focused most of its first 100 years on the organization of scientific symposia and science diplomacy following both world wars. Over the last decade, the IAU activities have broadened considerably to include communication with the general public, education and training, and using astronomy as a tool for development and capacity building, as outlined in its Strategic Plan 2020-2030 [2]. To implement its mission, the IAU has created four Offices: (a) the Office of Astronomy for Development (OAD), a joint venture with the South African National Research Foundation; (b) the Office for Astronomy Outreach (OAO), a joint venture with the National Astronomical Observatory of Japan; (c) the Office for Young Astronomers (OYA), a joint venture with the Norwegian Academy of Sciences and Letters; and (d) the Office of Astronomy for Education, a joint venture with the Haus der Astronomie in Heidelberg. Their efforts, together with activities by thousands of researchers and volunteers worldwide, are key to bringing astronomy's benefits to society.

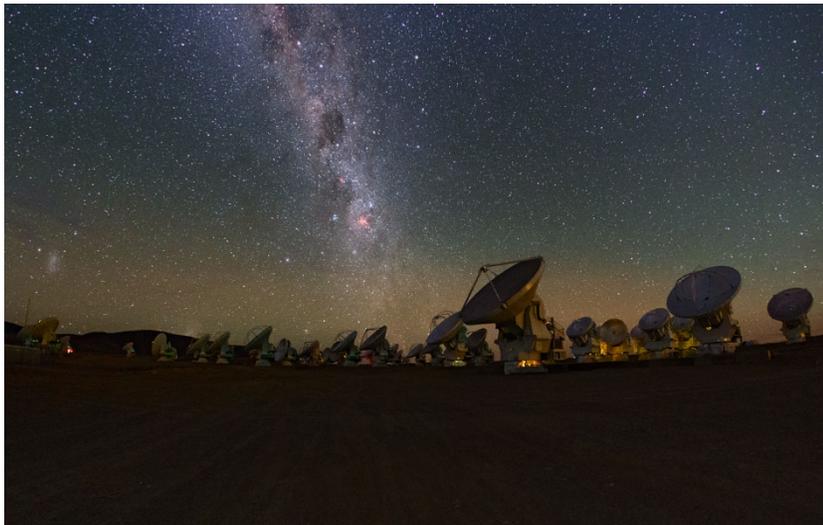

Milky Way seen over the Atacama Large Millimeter Array (ALMA), a worldwide collaboration between Europe, East Asia and North America. ALMA consists of 66 antennas on a high-altitude plane in northern Chile (credit: ESO/B. Tafreshi).

Another priority goal of the IAU is to protect the dark and (radio) quiet skies, which are being threatened by huge increases in urban lighting, radio interference and the launch of satellite constellations. These developments affect not just astronomy, but also human health, wildlife and the fundamental right of every human to view the night sky. International collaboration and agreements will be key to mitigating the effects.

The IAU also serves as the internationally recognized authority for assigning designations to celestial bodies and the surface features on them. For example, Asteroid 14429 Coyne was named in honor of Father George Coyne, who was Director of Vatican Observatory from 1978-2006.

**Inspiring the younger generation**

To address big challenges like COVID-19 or climate change, society needs bright and curious young minds that are trained in scientific methods. Astronomy provides an exciting low-threshold gateway for young children into the sciences. Questions on what constitutes black holes, what causes the most powerful explosions in space, and what is the possibility for life elsewhere in the Universe are among those that inspire them most [3]. Astronomy also sparks their curiosity, which is ultimately a major driver for innovation.  Once inspired, they are as teenagers more likely to choose a career in science and technology. This holds equally well for girls and boys. Successful programs like 'She speaks science' [4] use the power of storytelling to promote women and minority scientists and attract young girls to STE(A)M under the motto: 'the world needs science and science needs more women'.

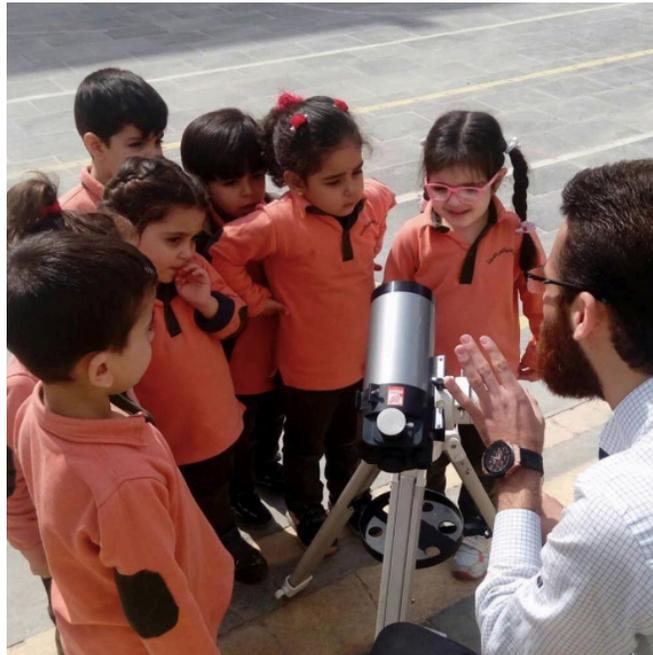
IAU-OAO outreach event. Source: https://www.iau.org/public/oao/oao_global_events/

**Technological advances**

As the oldest of the sciences, astronomy has played a fundamental role in the development of the human race. The ability to predict the motions of the Sun and stars was a decisive factor in the emergence of agriculture, time keeping and navigation in the earliest civilizations. The drive to detect faint astronomical signals and conduct large sky surveys has continued to spur technological advances, such as in detector development and computational power.

Instrumental techniques and algorithms used in astronomy find their way into medical, ICT (information and communications technology) and environmental research, although such technological spin-offs may take 20-30 years to develop after discovery [5]. Wi-Fi is one of the most important technologies to come out of astronomy. Other examples include GPS systems requiring precision time-keeping and applications of relativity, adaptive-optics laser eye surgery, medical imaging including MRI and CAT scans, traffic flow simulations, grid computing, measuring atmospheric particulates and skin cancer diagnosis techniques. Large scale challenges such as global climate change, energy production and understanding life itself directly benefit from a better understanding of (the atmospheres of) planets, both within and outside our Solar System, and of the nuclear fusion processes that occur in stellar systems. Especially now, new technological devices and the ability to handle and analyze huge data streams are important factors in tackling pandemics like COVID-19.

**Education**

Astronomy-trained students need to employ a 'system-level' approach to research, a skill that is also essential for addressing major societal challenges. They acquire strong foundations in mathematics, physics, chemistry, computation and statistics, and instrumentation. Astronomical research is not addressed by isolated table-top experiments, but derives from observing a diverse, distant and uncontrolled environment.  To tackle highly complex astronomical questions, astronomers are required to develop scenarios, design and build new instruments, cope with practical limitations of real data, develop algorithms for handling data imperfections, plan further observations to test models, and carry out massive numerical simulations.  These steps force students to think broadly about how to solve complex problems using a wide range of tools and in collaboration with people with different technical backgrounds. Working in international teams also hones communication skills, cultural exchanges and an awareness of cultural diversity and differing perspectives [6].

**Using astronomy for development**

Astronomy is an excellent pathway for capacity building, since students are trained in high-tech and big data analysis [7].  This combination of skills makes astronomy students attractive to a wide range of sectors in society, ranging from consultancy firms, ministries, ICT, industrial labs, health care, and banks. Astronomy is therefore a human capital developer whose impact far exceeds that of cutting-edge research alone. Indeed, developed and developing countries alike have recognized that investments and advances in astronomy, space science and technology contribute to their economic growth, development, and well-being [8,9, 10]. Science, far from being merely a luxury, is at the heart of society.

**Astronomy and COVID-19**

Given the broad applications of astronomy to other areas, it is not so surprising that astronomers can apply their expertise to assist in the COVID-19 pandemic in many ways. For example, astronomers and engineers with expertise in gas-handling equipment used in astronomical research are working together in several countries to design and manufacture ventilators (including those requiring no electricity) for COVID-19 patients, as well as decontamination systems and personal protective equipment (PPE) [11]. Astronomers with supercomputer access are helping to model the structure of the virus and its spread.

To help ameliorate some of the negative consequences of the current pandemic, the IAU-OAD has fast-tracked a call for proposals that use astronomy in any of its aspects. Some of the several dozen funded projects directly related to COVID-19 include a visualization dashboard providing information on healthcare facilities in South Africa,  providing hygiene support in Nigeria, mapping COVID-19 hotspots, producing and distributing PPE, disseminating information about COVID-19, distribution of food, and organizing an epidemiologists-astronomers hackathon to include appropriate population characteristics in epidemiological models. Several other projects are aimed at sharing astronomy resources through online data and access to remote observing, physically distributed "astronomy in a box" toolkits, webinar training workshops for educators as well as researchers, online lectures, and online tutoring. Projects to help lift people's spirits during these stressful times include stargazing to appreciate our shared heritage of the night sky, art inspired by astronomy, and video essays [12].

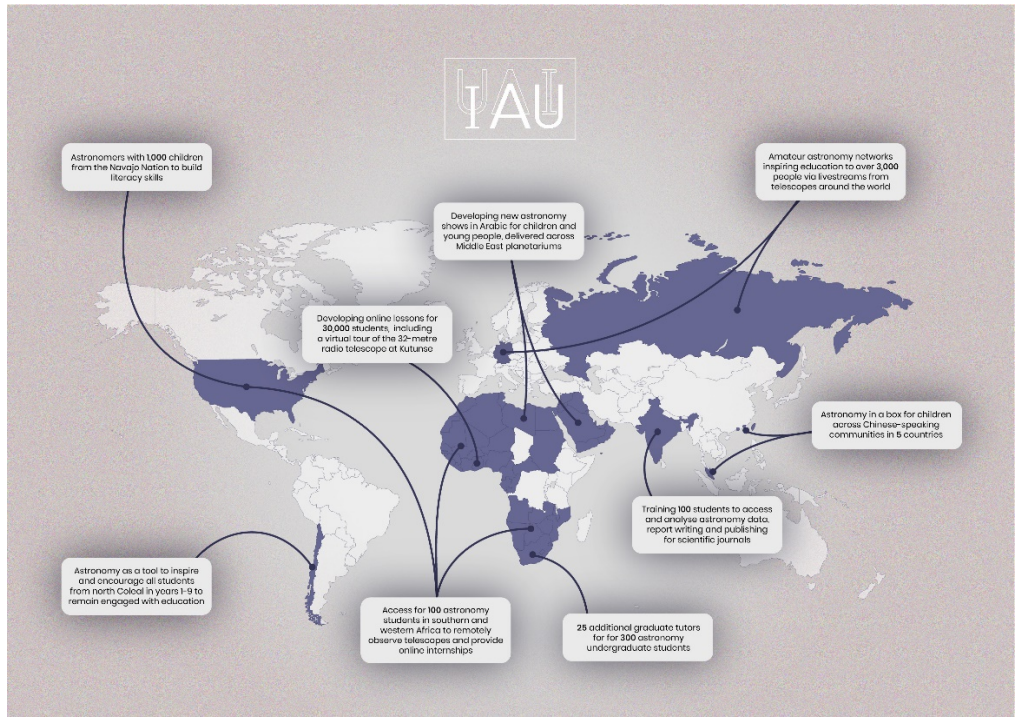

Some IAU projects to mitigate the effects of the pandemic. Credit: Aneta Margraf-Druc. Source: https://www.iau.org/news/announcements/detail/ann20030/

Other OAD projects include assistance to teachers to deliver their STEM lessons online; purchase mobile data for people to access information on the pandemic as well as to use for schooling; and supply hard copy materials for students without internet access. The IAU-OAO has also compiled a large number of digital activities to carry out at home, using the natural laboratory of the sky to motivate and educate children and parents alike [12]. OAD and OAO together have funded initiatives to develop and distribute educational materials to reach populations with little or no internet access.

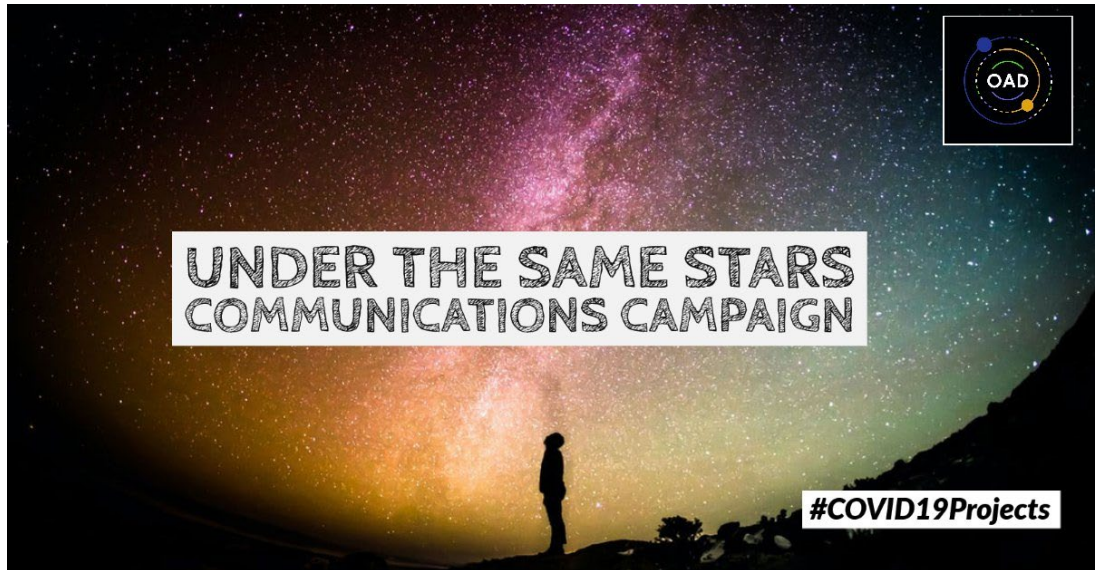

A nighttime viewing project to relieve stress during the pandemic.
Source: https://twitter.com/Astro4Dev/status/1305401093703204864/photo/1

**Perspective**

The discovery of exoplanets - planets around other stars than the Sun - has demonstrated that Earth and our solar system are not unique. In fact, planets outnumber the several hundred billion stars in the Milky Way. We now know that we live on a small rocky planet circling an ordinary star in the outskirts of our galaxy, hardly anything or anyplace special. The next quest in astronomy is to determine whether those other planets could be habitable and show signs of life by studying their atmospheres. If Earth proves to be unique as a habitable planet, this may strongly boost humankind's efforts to preserve our planet. On the other hand, demonstrating that we are not alone would have an equally fundamental impact on philosophy, sociology, the arts, and all major religions.

Exactly 30 years ago on Valentine's day this year, the iconic Voyager-1 picture was taken, looking back at Earth, the "pale blue dot," from the outer regions of our solar system [14]. More than any other picture, it drives home the message that our planet is small and fragile in space. This is the perspective that only astronomy can bring, inspiring awe while eliciting humility and tolerance. Ultimately, we are all citizens under the same starry sky.

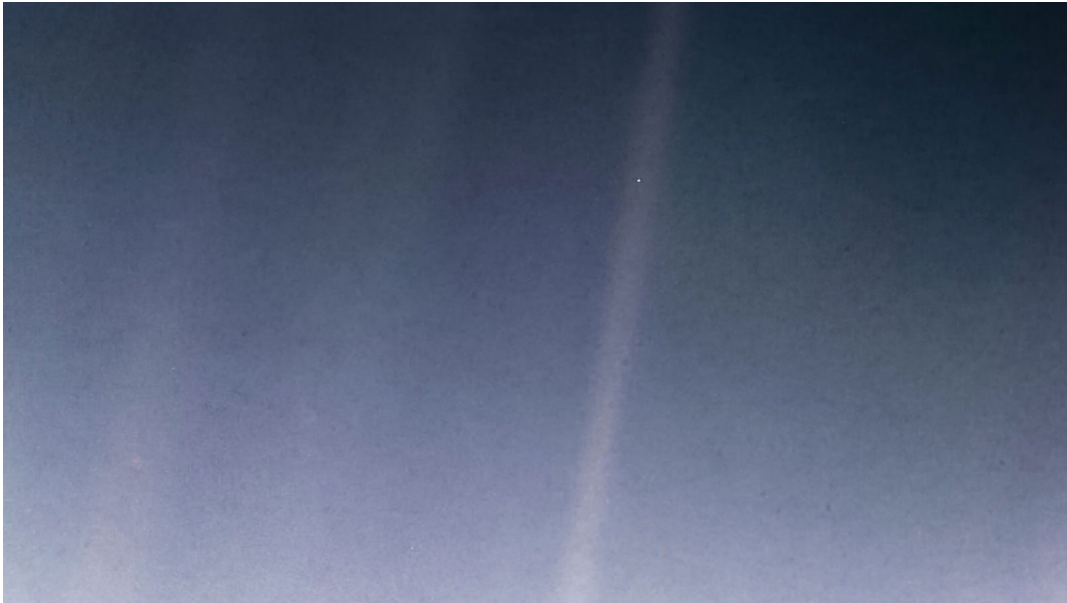
Earth from Voyager-1: the pale blue dot. Source: https://solarsystem.nasa.gov/resources/536/voyager-1s-pale-blue-dot/

**References**

[1] Martin Rees, On the future prospects for humanity, 2018, Princeton University Press
[2] IAU Strategic Plan 2020-2030, https://www.iau.org/administration/about/strategic_plan/
[3] S. Sjøberg et al. Relevance of Science Education (ROSE) study 2010, University of Oslo
     https://www.roseproject.no/publications/english-pub.html
[4] She speaks science, http://www.shespeaksscience.com/
[5] From Medicine to Wi-Fi: Technical applications of Astronomy to Society, IAU 2019,
https://www.iau.org/static/archives/announcements/pdf/ann19022a.pdf
[6] NOVA Vision 2025, Netherlands Research School for Astronomy, https://nova-astronomy.nl/wp-content/uploads/2016/10/VISION-2025.pdf
[7] IAU Strategic Plan 2010-2020, Astronomy for Development,
https://www.iau.org/static/education/strategicplan_2010-2020.pdf
[8] Khotso Mokele, "Using Astronomy to shape a country's science and technology landscape,"
Highlights of Astronomy, Volume 16XXVIIIth IAU General Assembly, August 2012, T. Montmerle, ed.
[9] Solomon Tessema in IAU Catalyst 3 June 2020, p. 25
     https://www.iau.org/static/publications/iau-catalyst-03.pdf
[10] Norman Augustine, National Academy of Sciences, National Academy of Engineering, and Institute of Medicine. 2007. *Rising Above the Gathering Storm: Energizing and Employing America for a Brighter Economic Future*. Washington, DC: The National Academies Press. https://doi.org/10.17226/11463
[11] http://www.astro4dev.org/role-of-astronomy-in-the-fight-against-the-covid-19-pandemic/
[12] https://www.iau.org/news/announcements/detail/ann20028/
[13] https://www.iau.org/public/callforonlineresources/
[14] https://solarsystem.nasa.gov/resources/536/voyager-1s-pale-blue-dot/